# Mapping the Design Space of Human-AI Interaction in Text Summarization

**Ruijia Cheng** [*]
University of Washington
rcheng6@uw.edu

**Alison Smith-Renner**
Dataminr Inc.
arenner@dataminr.com

**Ke Zhang**
Dataminr Inc.
kzhang@dataminr.com

**Joel R. Tetreault**
Dataminr Inc.
jtetreault@dataminr.com

**Alejandro Jaimes**
Dataminr Inc.
ajaimes@dataminr.com

## Abstract

Automatic text summarization systems commonly involve humans for preparing data or evaluating model performance, yet, there lacks a systematic understanding of humans' roles, experience, and needs when *interacting* with or being *assisted* by AI. From a human-centered perspective, we map the design opportunities and considerations for human-AI interaction in text summarization and broader text generation tasks. We first conducted a systematic literature review of 70 papers, developing a taxonomy of five interactions in AI-assisted text generation and relevant design dimensions. We designed text summarization prototypes for each interaction. We then interviewed 16 users, aided by the prototypes, to understand their expectations, experience, and needs regarding efficiency, control, and trust with AI in text summarization and propose design considerations accordingly.

## 1 Introduction

In this era of rapid information consumption, access to high-quality summaries, such as online news highlights and research paper abstracts, is increasingly important. However, summarization is difficult for humans, demanding high cognitive load and expertise (Hidi and Anderson, 1986). Algorithmic approaches can automate summarization but typically underperform humans and require many high-quality human-written summaries for training (Durrett et al., 2016; Huang et al., 2021). AI systems that involve humans typically constrain their input to data preparation (Lloret et al., 2013) or final evaluation (Khashabi et al., 2021) as the first or last step in the summarization workflow. How can humans work together with AI to produce better summaries?

Other text generation tasks, such as machine translation and creative writing, offer inspiration.

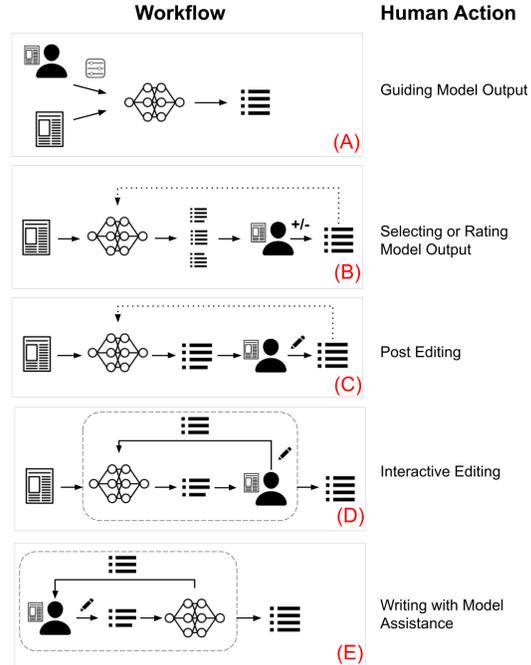

Figure 1: Five human-AI interactions in text generation from Study 1, illustrated as summarization tasks. Explanation of the actions and visual elements are in §2.2.

Beyond being data providers and evaluators, humans can generate text with the assistance of AI. For example, humans can revise AI translation (i.e., "post-editing") (Green et al., 2013) or write with AI suggestions (Clark et al., 2018). These interaction techniques may apply to summarization, yet our understanding of the possible human-AI interactions and how to design for them is incomplete.

To that end, we conducted two studies to explore: **RQ1** What are the different ways that humans and AI can interact in text generation? **RQ2** What are humans' experience and needs with these different types of AI assistance in the context of summarization? For these studies, we took a human-centered approach—focusing on what humans want from AI and how to improve their experience, rather than improving AI models. In Study 1, we conducted a systematic literature review of 70 papers that involve AI-assisted text generation systems and developed a taxonomy of five different human-AI interactions (Figure 1), including distinct human actions, controls, workflows, and interface features. In Study 2, to explore human experience with these five interactions in the specific context of text summarization, we conducted interviews with 16 users using prototype interfaces that we developed and identified varied user needs regarding efficiency, control, and trust, informing design considerations for AI-assisted text summarization and generation systems.

---

[*] Research work was done while the author was interning at Dataminr Inc.

This work contributes: 1) the first known systematic literature review that provides a taxonomy of human-AI interactions for text generation, 2) an interview user study to understand user experience and needs in each interaction, and 3) an outline of the design space for AI-assisted text summarization and broader text generation systems. This research is a formative, initial exploration that first maps the different types of human-AI interaction for text generation and then surfaces user needs and perceptions for each type in the context of human-AI summarization. Such formative work is crucial for understanding what users might want and need from systems without biasing them by existing implementations—future researchers can use these findings to design and evaluate new human-AI text summarization systems.

## 2 Study 1: Systematic Literature Review of Human-AI Text Generation

For the first study, We performed a systematic literature review of human-AI text generation and developed a taxonomy of existing interaction types, which differ by the level of human control over the output, whether the AI iteratively updates its output based on human interaction, and whether human or AI initiates the interaction workflow. This study identifies and synthesizes the types of interaction explored in prior work. While we do not contribute any entirely new human-AI interactions for text generation, our taxonomy provides grounding for our second study—exploring user needs and experience with these interactions—as well as future research and design in this space. Future AI/HCI researchers can refer to this taxonomy when designing text generation systems or experiments to select the interaction and design elements that best fit their scenario and to compare findings with prior work toward a more formalized understanding of the space.

### 2.1 Method

We conducted a comprehensive search on academic papers about human-AI interaction in text generation from online indices and relevant workshops, e.g., ACM Digital Library,[1] arxiv.org,[2] HCI-NLP.[3] With the 692 papers queried from these sources, we manually coded the titles and abstracts against our inclusion criteria (e.g., for the goal of text generation, engaging humans beyond data preparation or evaluation) and identified a final set of 70 papers that describe human-AI interaction (or AI assistance) for the humans' goal of generating text (opposed to offline model training or evaluation). See further details on sources, query methods, inclusion criteria, and analysis process in Appendix A.1.

### 2.2 Findings: Five Human-AI Interactions in Text Generation

From the 70 papers, we identified five human-AI text generation workflows featuring distinct human actions—guiding model output, selecting or rating model output, post-editing, interactive editing, and writing with model assistance (Figure 1). We present the taxonomy that summarizes our findings in Table 1. For each interaction, we describe the human action, the type of human control, whether the model iterates based on human action, who (human or model) initiates the workflow, and interfaces from literature. We explain the interactions and visual elements in Figure 1 as follows:

**Guiding Model Output.** Humans can give model guidance to generate text (Figure 1, A). Here, humans are in the driver seat of the generation process with power to initiate and control the final output. Humans provide preferences (illustrated as the slider icon in the figure) to the model. The model takes human input and original text (the document icon) and produces text as the final product (the rightmost four-line summary icon), concluding the interaction with no further iteration. The specific guidance that a human can provide is varied and depends on their ML expertise. Model developers can adjust model parameters such as neural attention weights and hyperparameters. Alternatively, lay-users can prompt the model using semantic cues (e.g., keywords, topic tags, or descriptive prompts), or specify the style or details of the text (e.g., sentiment or length). Some interfaces support selecting or deselecting what from the input text should be used by the model.

---

[1] https://dl.acm.org/
[2] https://arxiv.org/
[3] https://aclanthology.org/2021.hcinlp-1.0/

| Human Action | Human Control | Model Iteration | Workflow Initiation | Interfaces |
| --- | --- | --- | --- | --- |
| Guiding model output | Generating | No | Human | **Model adjustment (e.g., attention weights and hyperparameters):** (Zhang et al., 2011; Passali et al., 2021); **Semantic prompts (e.g., keywords, topic tags, and written prompts):** (Pouliquen et al., 2011; Ghazvininejad et al., 2017; Clark et al., 2018; Fan et al., 2018; Zarinbal et al., 2019; He et al., 2020; Osone et al., 2021; Chang et al., 2021; Strobelt et al., 2021); **Style specification (e.g., sentiment slider, length of text):** (Ghazvininejad et al., 2017; Freiknecht and Effelsberg, 2020); **Modification on input (e.g., select/deselect chunks of input text):** (Pouliquen et al., 2011; Zhang et al., 2011; Gehrmann et al., 2019) |
| Selecting or Rating Model Output | Selecting (or rating) | No | Model | **Multiple choices:** (Zhang et al., 2011; Pouliquen et al., 2011; Kreutzer et al., 2018; Rosa et al., 2020; Stiennon et al., 2020); **Rating (e.g., Likert scale, numerical, binary):** (Nguyen et al., 2017; Lam et al., 2018, 2019; Zarinbal et al., 2019; Bohn and Ling, 2021; Wu et al., 2021) |
| Post-editing | Editing | No | Model | **Free-form text box:** (Denkowski and Lavie, 2012; Yao et al., 2012; Yamaguchi et al., 2013; Turchi et al., 2014; Turner et al., 2015; Chu and Komlodi, 2017; Huang et al., 2020; Moramarco et al., 2021); **Suggestion for edits (e.g., chatbot, substitution dropdowns):** (Liu et al., 2011; Shi et al., 2013; Weisz et al., 2021; Passali et al., 2021); **Scaffold for context (e.g., embedded dictionary):** (Sugiyama et al., 2011; Lin, 2011); **Productivity support (e.g., editing priority, auto-correction):** (Lagarda et al., 2015; Peris and Casacuberta, 2019b; Wang et al., 2020; Zhao et al., 2020; Weisz et al., 2021) |
| Interactive Editing | Editing | Yes | Model | **Prefix-based edits:** (González-Rubio et al., 2013; Peris and Casacuberta, 2018, 2019b,a); **Edits at arbitrary locations:** (González-Rubio et al., 2016; Weng et al., 2019) |
| Writing with Model Assistance | Generating & Editing | Yes | Human | **Auto-completion:** (Green et al., 2014; Torregrosa et al., 2014; Gero and Chilton, 2019; Santy et al., 2019; Buschek et al., 2021; Ippolito et al., 2019; Calderwood et al., 2020; Bhat et al., 2021; Clark and Smith, 2021); **Substitution dropdowns:** (Green et al., 2014; Torregrosa et al., 2014; Santy et al., 2019; Ippolito et al., 2019; Gero and Chilton, 2019; Calderwood et al., 2020; Buschek et al., 2021; Padmakumar and He, 2021); **Asynchronous suggestions:** (Torregrosa Rivero et al., 2017; Clark et al., 2018) |

Table 1: A taxonomy of human-AI interaction for text generation: human action, the type of control that humans have over the final output (Human Control), including generating, editing, or selecting/rating AI-generated text, whether the AI iterates, dynamically providing updated outputs based on human interaction (Model Iteration), who (Human or Model) initiates the workflow (Workflow Initiation), and what interfaces are used (Interfaces).

**Selecting or Rating Model Output.** Humans can select from or rate generated texts (Figure 1, B). The model initiates such workflows by generating candidates of final output (the 3 four-line summary icons). The human does not directly generate text, instead, they have the control to select from or rate the candidates (the "+/-" icon) to support final output. We focus on the case where human input decides which candidate is chosen as the final product, but such feedback can also be used for online model training (e.g., active learning), represented as the dotted arrow going from the final product to the model in the figure.

**Post-editing.** Post-editing (Figure 1, C)—common in machine translation (Green et al., 2013)—starts with text drafted by AI (the four-line summary icon with an incomplete last line), which humans edit (the pen icon). The workflow concludes when the human finishes editing, without any further iteration by the model, although the edited text can be used for future model training (the dotted arrow going from the final product summary back to the model). In many interfaces, users post-edit AI-generated text in text boxes. Some systems include innovative editing paradigms (e.g., chatbots) and other supports, such as drop-down menus for word or sentence substitution and scaffolds for context (e.g., embedded dictionaries) to aid understanding. Further, productivity supports can reduce workload, such as signalling where edits are needed and automatic error amendment based on users' editing history.

**Interactive Editing.** Humans can edit text interactively with AI (Figure 1 D). Like post-editing, this dynamic editing interaction is also commonly used in machine translation systems, i.e., "interactive machine translation" (Barrachina et al., 2009). First, the model generates an incomplete draft (the three-line summary icon with a incomplete last line), to which human provides edits. While post-editing would stop at this point, in interactive editing, the model iterates on the human-edited text (the complete three-line summary on the top) to update and generate more text for continued human editing, iteratively and in real-time (the solid line arrow). This iterative human-AI interaction (in the dashed frame) results in the final product. Interactive editing interfaces take many forms. "Prefix-based" edits are specified on a left-to-right phrase by phrase fashion, while the model makes new predictions on the rest of the sentence. In other interfaces, humans edit at arbitrary positions of the AI-generated sentence and the model updates the whole sentence. Some systems offer additional editing support, such as highlighting necessary edits and dropdowns for substitution.

**Writing with Model Assistance.** Finally, humans can write with AI assistance (Figure 1, E). Humans initiate this workflow and have a high-level control: humans begin writing while the model provides suggestions and can revise their writing based on the suggestions or ignore them. This iterative human-AI interaction (in the dashed frame) generates the final text. In this process, the model iterates, providing additional suggestions based on humans' writing. Assisted writing interfaces include real-time auto-completion, which can happen at the word, phrase, or sentence level. Some systems offer alternative suggestions in dropdowns from which humans choose. Others provide asynchronous suggestions, presented after users have finished writing to reduce distraction.

## 3 Study 2: Interview Study on AI-assisted Text Summarization

We present a user study in which we evaluated interactions in AI-assisted text summarization with the context of text summarization through interviews aided by prototype interfaces. Our goal is not to prescribe which is "best" but to achieve a qualitative understanding of user needs with each interaction to inform future research and design.

### 3.1 Prototype Design

We first developed prototype interfaces to represent the five interactions identified in Study 1. We used these prototypes in our user interviews to elicit needs, expectations, and experience around AI-assisted text summarization.

While some prototypes for these interactions exist in the literature for broader text generation tasks, many include additional features and visual design that may affect users' perceptions, therefore, we develop our own set of consistent, simple prototypes for exploring text summarization specifically. Each interactive prototype, implemented in Figma[4] or Google Docs, allowed participants to read an online news document and generate summaries with the support of a hypothetical AI model.

All prototypes were built based on the "Wizard-of-oz" prototyping concept (Kelley, 1983), commonly used in user studies on intelligent systems. This concept allows users to interact with intelligent systems that are not fully implemented; instead, "system outputs" are prepared manually by the research team. This method allows designers of intelligent systems to rapidly test design concepts, understand user experience, and iterate on their design. This method has been used to prototype NLP systems, for example, in the design of chatbots (Zhou et al., 2019; Avula, 2018).

In our study, participants interacted with "Wizard-of-oz" prototypes instead of implemented AI models, so that we could explore human perspectives and user needs for the different interaction types without being limited by model performance or other system characteristics, such as unpredictability and slow updates. Each of our five prototypes used the same original text (a news article from the set used in the warm-up activity, explained in §3.2) that needed to be summarized. The prototypes differed by the interaction interface they supported (from Study 1). Depending on the interaction type, the prototype presented users with "AI-generated" summaries, outputs or suggestions that were pre-defined and written by the research team. While the human-written outputs used in this study might not necessarily imitate the quality and style of AI-generated summaries, outputs, or suggestions, they were intended to provide concrete examples of the interaction types and elicit participant's needs and expectations. Future work should explore how more realistic model outputs affect human perceptions. The specific design of each prototype (and screenshots) are described in the Appendix A.3.

---
[4]https://www.figma.com/

### 3.2 Method

**Participants.** We recruited 16 participants (10 females, 6 males, all based in the U.S.) from Upwork[5] with varied professional backgrounds and varied familiarity with the domains of Reddit posts, online news, and U.S. government bills. We intentionally recruited participants who have at least some level of familiarity with text editing or summarization from professional or educational settings, which ensures that participants could speak about for which summarization tasks they desire assistance and describe their needs for such interactions. See Appendix A.2 for demographic details. The study took 2.5 hours, and we paid each participant $60.[6]

**Procedure.** Each participant first did a 60-minute offline warm-up activity less than 48 hours before the interview, where they summarized six articles (two Reddit posts on scams or finance,[7] two news articles from CNN/Daily Mail,[8] and two U.S. government bills[9]). This activity aimed to expose participants to summarization with articles written in different styles and with varied domain context.

Then, during the 1-on-1 semi-structured recorded video interviews (90 minutes), participants first reflected on their experience in the warm-up summarization tasks and then interacted with all five prototypes, in random order, as users of AI-assisted summarization systems. Participants were shown a news article from the warm-up task (all prototypes used the same article) and also asked to imagine using the prototypes for the other documents from the warm-up. They interacted with the interfaces and received pre-determined outputs that mimicked AI assistance:

1. **Guiding Model Output**: participants could change the desired summary length and style (formal or informal) using sliders and highlight parts of the original text that should be in the summary. We asked participants what additional guidance they wanted to offer to the model.

2. **Selecting or Rating Model Output**: participants chose from three AI-generated summaries.

3. **Post-editing**: participants saw an editable AI-generated summary (text box) and talked through how they would edit it.

4. **Interactive Editing**: given an AI-generated summary (text box), participants chose possible edits to the first sentence (dropdown menu) and then requested the model to update the summary based on those edits. We asked participants to imagine an alternate interface where they could edit anywhere in the summary.

5. **Writing with Model Assistance**: following a "wizard-of-oz" prototyping method (Kelley, 1983), a researcher acted as an AI bot in a Google Doc. As the participants typed their summaries, the "bot" provided suggested next sentences and added comments.

Participants were then prompted to talk through experience with each prototype, including what they liked or disliked regarding efficiency, control, and trust, and how they would improve them. See Appendix A.3 for screenshots of each prototype and A.5 for the interview questions and instructions used in the study. We collected and transcribed 22.6 hours of interview recordings, which were analyzed using thematic analysis (Guest et al., 2011). We performed two rounds of open coding and developed themes reported in the following sections. We refer to participants as P1-16 with gender non-specific pronouns (i.e., they, them). We present findings regarding efficiency, control, and trust—common themes in our interviews and key dimensions in the human-AI interaction literature (Amershi et al., 2019; Shneiderman, 2020). We include additional findings about user expectation and needs in writing summaries in Appendix A.4.

### 3.3 Findings

#### 3.3.1 General Expectations & Needs

**Expectation on AI to improve summarization.** The warm-up summarization tasks were challenging and tedious. Summarizing the Government Bills was slow for many due to the unfamiliar domain, jargon, and *"super dry, super repetitive"* (P3) style. Summarizing informally-written and opin-ionated Reddit posts was also difficult as many were unsure about whether to keep the authors' perspective or summarize neutrally. Therefore, participants hoped that AI could speed up and ease this process. Many envisioned AI-generated summaries as a useful starting point: *"determining where to get started can be a big roadblock for some writers...being able to have that auto-generated summary as your baseline to develop your ideas off I think would be really helpful"* (P13) Further, participants hope to use AI suggestions to improve the content and structure of their writing: *"it gives more of a third party look at things...just kind of compare and contrast it to what I've done, to make sure that I'm on the right track"* (P15).

---

[5] https://www.upwork.com/
[6] Adequate payment in the United States.
[7] Extracted from r/scam and r/wallstreetbets
[8] https://paperswithcode.com/sota/text-summarization-on-cnn-daily-mail-2
[9] https://www.tensorflow.org/datasets/catalog/billsum

**Different desire for control.** Most agreed they at least wanted the ability to proofread the AI-generated summaries, or to *"have the final say"* on whether it was good as a final product (P3). Some said this *responsibility* was a habit of professionalism; others were cautious of the work done by AI and wanted to ensure quality: *"it was drilled into my head that these devices are tools and they can fail...we're always responsible for overseeing what the computer does"* (P7). Beyond simple editing, participants had a varied desire for control. Some felt summarization was *"not necessarily a creative enterprise"* and, therefore, were willing to *"relinquish a little bit of control to AI"* for efficiency (P3), while others wanted to participate in the entire generation process. These participants preferred to compose their own summary using AI strictly as an aid, e.g., *"it would just simply be used as a tool for me, not as something to replace my work"* (P12). Many felt uncomfortable using AI-generated summaries directly or after only proofreading edits due to the sense of *"plagiarism"*, and as a result, wanted the ability to rewrite summaries into their own words. Desired control could also vary by situation. For example, P7 wanted more control when summarizing for the bills because that was a more *"serious and important task,"* while P8 would be more lenient when summarizing the Reddit posts: *"even [the summary] doesn't capture everything, it is good as long as the summary kind of outlines the the key points of the article."*

**Need to understand AI to reduce over-reliance and boost trust.** Participants were concerned that they might rely on AI too much and lacked confidence to correct it even when it was wrong. For example, P7 felt AI-generated summaries were an *"authority that has given you this thing"*, saying that *"for most people, if presented with something, they're going to go with it."* As a result, users could lose confidence when they disagree with the AI. P8 shared their hesitation to dramatically edit AI-generated summaries: *"it's almost feeling like you're pivoting against the AI...should I question what the AI thinks is important?"* This apprehension might increase when participants are summarizing for unfamiliar or difficult documents. Specifically, some anticipated a lack of trust when summarizing challenging articles because they could not reasonably assess the AI's output: *"I probably wouldn't use it for a lengthier subject that I wasn't familiar with...just because I wouldn't know if the AI was writing something I wanted to write"* (P6).

To foster trust, many wanted information about *how* the AI generated the summaries or suggestions—why the certain information is included and whether there were any hidden presuppositions by the model. For example, P8 said, *"knowing, in a very basic sense, how the AI is generating these summaries, [will] give me a good idea of essentially how much I can trust it."* As the prototypes did not include explanation features, participants noted that they did not trust the AI since they did not understand the mechanism, as P12 put: *"there's too many variables that you don't know. Too many unknowns for me."*

### 3.3.2 Interaction-Specific Experience & Needs

We report participants' needs and expectations on efficiency, control, and trust when interacting with each of the five interfaces and present a conceptual comparison between the five interfaces in Figure 2.

**Guiding Model Output** Participants felt this prototype streamlined the summarization process, as they did not need to compose the summary themselves, but only give their preferences: as P6 said, *"I don't have to do quite as much thinking...I don't have to type which takes time."*

Most appreciated the control over the summary generation by adjusting parameters. For example, P8 liked the text highlighting feature, as *"it gives you the amount of control in terms of being able to choose the parts that you think are important."* Many envisioned using the interface to customize the summary for their target audiences. For instance, P6 imagined using it to tailor summary styles for different colleagues: *"with my staff, I*

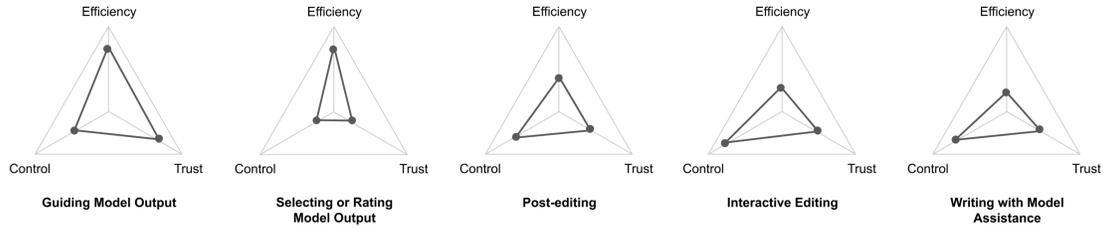

Figure 2: Illustration of participants' perception on level of efficiency, control, and trust with each interaction. These conceptual level charts show a qualitative, rather than precise, comparison between interactions.

*would use the short style...for my boss, I might use a longer formal summary to look a little more professional."* On the other hand, some were concerned about the lack of editing control: *"it doesn't have as much control as it seems. When you get to this [final] stage, you're stuck with it"* (P7).

Participants felt they had a reasonable understanding of the AI mechanism in this prototype and thus could trust it. Since they could change parameters (e.g., length, style) to experiment with different aspects of the summary, they better understood the process: *"we could [trust it more] maybe because I can play around with it. The long and short allows me to kind of have control"* (P12).

**Selecting or Rating on Model Output** Most participants thought this interface could make summarization much more efficient, as they only needed to choose the best summary out of a list rather than to write or edit. Participants valued this low workload: *"[it was] gonna summarize the article for me and do all the work...whatever I can use to buy out my part of the labor I am all for."* (P6)

Despite the efficiency advantage, many complained about the lack of control, specifically the inability to influence or edit the AI-generated summaries. For example, some felt that choosing the best might not ensure quality: *"what if three of these are presented, and none of them are really good enough. Then it's just a matter of picking the least bad one"* (P7). Further, since comparing and selecting were simpler tasks than writing or editing, participants paid less attention and thought less critically: *"evaluating already written summaries and trying to decide which one is the best is different from just writing your own summary...I am not like super mentally invested in it, if I were writing my own, I'd very careful with word choices"* (P5).

Many struggled in selection as they did not know how the candidates were generated: *"how do you determine, from an AI standpoint, what information to keep and what information to get rid of? How do you determine the priority as what stays and what goes? Is it biased in any way?"* (P14).

**Post-editing** Participants were mixed on how much efficiency post-editing would bring. Some thought it could make summarization faster: *"I could just run it through this, and then edit it and change the things that I needed to change. It would save me a lot of time and energy"* (P1). For others, editing unfamiliar text was an equally time consuming task. For instance, P8 would always ensure the summary aligns with their personal writing habits and style, and, therefore, they would spend a lot of effort in editing and customization: *"I feel like it would be just as much work to just write it from scratch...if I'm trying to make it original, I have one less way of being able to word it"* (P8).

Participants were satisfied with the editing control granted by this interface. For instance, P10 shared why they liked it more than the Selecting or Rating on Model Output interaction *"You can edit it to however you'd like. I think the freedom to edit appeals to me a little bit more."*

Similar to other prototypes, participants hoped to see more information on how the AI generated the summary. For example, P16 said it would be helpful to visually see which parts of the article were emphasized in the AI-generated summary, so that they could decide what to focus on. Similarly, P7 imagined a quantifiable way to indicate how much of the content in the summary was matched with the original article. They hoped for *"some advanced algorithms checking to make sure that it did it right"* to decide how much to trust it.

**Interactive Editing** Participants were more suspicious about this prototype's efficiency compared to the others. They worried that the dynamic updates to the summary might disrupt their existing edits, from P9: *"what if I rewrite the first sentence again, and then it changes everything else. I feel* *like it can really start to ripple.*" Real-time model updates also generate new text, creating more editing work: *"with every choice that I have written here, I'm going to have another choice down here to consider whether I want to use or not. It's gonna cause me way too much work"* (P12).

For control, many valued that in addition to editing, they could also experiment with different versions due to the dynamic updates. However, some viewed it pointless to iterate with the AI and would rather complete editing in a single turn: *"it's giving me a choice that I don't necessarily want...I want it to be as close to a final draft as possible, because then my editorial choices are final and have the feeling of finality"* (P3). Participants also worried about unpredictable AI actions that might impact their edits: *"I don't have any idea what the second paragraph is going to be until I make a choice with the wording of the first paragraph"* (P11); *"it is kinda stressful because if you use just one different word, it's going to change the entire thing"* (P15).

To ease this uncertainty, many wanted to understand how the AI updates based on their edits. P5 shared that they tended to discuss with coworkers on how certain choices were made—*"every word is intentional."* And, they hoped to have similar interaction with the AI, *"to know the reasoning behind the changes, the kind of logic flow,"* so that they could make better decisions on what to edit.

**Writing with Model Assistance** Participants felt that writing and iteratively making improvements based on new model suggestions and comments could be tedious: *"I have to go back and read what it suggests and see if it makes sense for them and for me. I just think it takes you more time to do this"* (P6). Also, from P14, it *"puts more back on the person writing the summary,"* which could introduce writer's block and stress.

Despite of the high control over the final output and whether to take AI's suggestions, participants wanted to control *when* they received assistance during summarization. Many viewed the auto-completion and suggestion intrusive and distracting, especially when they were not ready to receive help: *"it's harder to write when you have constant changes being thrown your way"* (P9). Comparing it with the Interactive Editing interface, P7 found the latter allowed more control over when AI helps: *"since you're pressing a button, you still feel like you have some control. And you have control the timing too, which is important, because, what if you want to think about your first sentence?"*

Similar to other interfaces, participants also wished to know why the AI made certain suggestions, so that they could decide whether and how to follow: *"I am a why person and I like to understand what I am doing. So if you're telling me to change something, you need to give me the reason"* (P6). In addition, some thought auto-completion might amplify human mistakes as it was learning from their writing: *"when I wrote my first sentence, I wasn't confident... And then for the bot to come in with that... it's not going to be a good summary, because I didn't know what I was writing."*

## 4 Discussion & Design Implications

In the taxonomy, we synthesized five human-AI interactions in AI-assisted text generation. Through interviews, we surfaced user experience and needs with each interaction technique to inform future research and design. Although our user studies focused on summarization specifically, we believe our insights can be used for designing broader text generation systems. We discuss both general and interaction-specific implications as follows.

**Offer the post-editing option regardless of interaction type.** In general, humans like to have the *"final say"* on AI-generated text. Even when participants' role was choosing model output, they still wanted the option to edit to ensure quality. As such, future human-AI text generation systems should provide editing options for the final output.

**Ensure customizable timing of AI-assistance in writing.** We found unsolicited auto-completion and suggestions could disrupt users' writing experience. Future AI-assisted writing systems should allow users to easily turn off or adjust any automatic functions to decide when to receive help from AI. For example, users can press a button on-demand, instead of getting automatic suggestions.

**Align interactive editing with user expectation.** While humans in general like editing support such as dictionary or substitution suggestion, participants were more skeptical about dynamic updates in the Interactive Editing case, as AI may make big changes when they intended to make minor edits. Therefore, they desired to adjust the *extent* of AI-predicted updates based on their intention. Echoing literature on predictable AI systems (Daronnat et al., 2021), future interactive editing systems should consider user expectations and empower users to preview AI actions. Systems could also model human editing intention, perhaps via action history like number and location of edits, and adapt AI actions accordingly to better serve user goals.

**Support tailoring text to different audiences.** Another use case for AI-assisted writing is to *tailor* generated text to different audiences. For example, Systems with Guiding Model Output interaction allow humans to specify desired style, audience and use cases and generate customized text. Post editing and interactive editing systems can incorporate scaffolds that provide wording and format suggestions tailored for different scenarios.

**Address trade-offs between efficiency and control.** Guiding models or selecting model outputs are efficient actions, but humans have concerns about the lack of editing freedom and ownership on AI-generated text. Systems that leverage human editing power or support human writing with AI grant more flexibility, but require more effort. Users' needs in efficiency and control vary based on their goals and context. For summarization, participants felt more responsibility and thus desired more control on editing and generation when working on professional tasks or on texts that were perceived to be important (e.g., Government Bills). Future systems should consider these differences and assign different level of controls to users accordingly. For instance, systems can classify tasks based on importance and automate more when generating texts that are less important, while inviting humans to participate more in the generation and editing process for more important tasks.

**Foster appropriate trust on AI.** Our findings echo literature that humans can both over- and under-rely on AI systems (Bussone et al., 2015; Buçinca et al., 2021). For example, consistent with Bhat et al. (2021), users may view AI-generated text as an authority and be conservative on making edits. Others were uncertain if the AI-generated text or suggestion is reliable, especially when working with important text. Our findings point to the need for *appropriate* trust on AI text generation in general. First, systems should support users to understand *how* the model generates text, so that they can decide whether to rely on it or not. One technique is to allow humans to participate in the model decision process. Systems can refer to the Guiding Model Output interaction and allow users to specify preferences and experiment with different outputs. Systems can also offer explanations to model mechanism, perhaps through visual representations (Zhang et al., 2011; Gehrmann et al., 2019). Second, systems should provide *context support*. Interview participants, regardless of interaction case, had issues working with AI when summarizing Government Bills, as they were unfamiliar with the format and jargon. To this end, systems should equip users with sufficient context, so that they can effectively evaluate AI suggestions and make decisions accordingly. For example, systems can offer embedded dictionaries, resource search, or user Q&A support.

**Limitations.** We note a few important limitations in our study. First, we scoped our paper specifically to human-AI interactions for the goal of generating text, while we did not study the needs of humans who develop models, annotate data, or consume final outputs from AI. Second, while we performed a formal systematic literature review in Study 1, we may have missed some important papers due to our sampling strategies. Therefore, our taxonomy of human-AI interaction for text generation might not cover all possible interaction types. Third, our interviews were limited in that participants performed only short-term interaction with a hypothetical AI model in predefined scenarios. Users may report different experiences when interacting with real AI models in real-world settings for longer periods of time. That said, our study is a formative first step that aims to ground future research. Future researchers and developers could design human-AI text summarization systems based on our findings and further evaluate the systems in more realistic settings, with realistic model outputs, and through large-scale experiments.

## 5 Conclusion

While humans are commonly asked to generate training data or evaluate final model output in text summarization, we draw attention to the potential of collaborative human interaction when working with AI. Our study first contributes a taxonomy of five types of human-AI interactions for text generation tasks. We provide insights on user experience and needs around efficiency, control, and trust for the five interactions and design implications, outlining a variety of considerations for researchers, developers, and designers working toward incorporating human users in text generation systems.

Linguistics, Vancouver, Canada, 43–48. https://aclanthology.org/P17-4008

Jesús González-Rubio, Daniel Ortiz-Martínez, José-Miguel Benedí, and Francisco Casacuberta. 2013. Interactive machine translation using hierarchical translation models. In *Proceedings of the 2013 Conference on Empirical Methods in Natural Language Processing*. 244–254.

Jesús González-Rubio, Daniel Ortiz-Martínez, Francisco Casacuberta, and José-Miguel Benedí. 2016. Beyond prefix-based interactive translation prediction. In *Proceedings of The 20th SIGNLL Conference on Computational Natural Language Learning*. 198–207.

Spence Green, Jeffrey Heer, and Christopher D. Manning. 2013. The Efficacy of Human Post-Editing for Language Translation. In *Proceedings of the SIGCHI Conference on Human Factors in Computing Systems* (Paris, France) *(CHI '13)*. Association for Computing Machinery, New York, NY, USA, 439–448. https://doi.org/10.1145/2470654.2470718

Spence Green, Sida I. Wang, Jason Chuang, Jeffrey Heer, Sebastian Schuster, and Christopher D. Manning. 2014. Human Effort and Machine Learnability in Computer Aided Translation. In *Proceedings of the 2014 Conference on Empirical Methods in Natural Language Processing (EMNLP)*. Association for Computational Linguistics, Doha, Qatar, 1225–1236. https://doi.org/10.3115/v1/D14-1130

Greg Guest, Kathleen M MacQueen, and Emily E Namey. 2011. *Applied thematic analysis*. Sage Publications.

Junxian He, Wojciech Kryściński, Bryan McCann, Nazneen Rajani, and Caiming Xiong. 2020. Ctrlsum: Towards generic controllable text summarization. *arXiv preprint arXiv:2012.04281* (2020).

Suzanne Hidi and Valerie Anderson. 1986. Producing Written Summaries: Task Demands, Cognitive Operations, and Implications for Instruction. *Review of Educational Research* 56, 4 (1986), 473–493. http://www.jstor.org/stable/1170342

Cheng-Zhi Anna Huang, Hendrik Vincent Koops, Ed Newton-Rex, Monica Dinculescu, and Carrie J. Cai. 2020. AI Song Contest: Human-AI Co-Creation in Songwriting. *ArXiv* abs/2010.05388 (2020).

Yichong Huang, Xiachong Feng, Xiaocheng Feng, and Bing Qin. 2021. The Factual Inconsistency Problem in Abstractive Text Summarization: A Survey. *arXiv preprint arXiv:2104.14839* (2021).

Daphne Ippolito, David Grangier, Chris Callison-Burch, and Douglas Eck. 2019. Unsupervised Hierarchical Story Infilling. In *Proceedings of the First Workshop on Narrative Understanding*. Association for Computational Linguistics, Minneapolis, Minnesota, 37–43. https://doi.org/10.18653/v1/W19-2405

John F Kelley. 1983. An empirical methodology for writing user-friendly natural language computer applications. In *Proceedings of the SIGCHI conference on Human Factors in Computing Systems*. 193–196.

Daniel Khashabi, Gabriel Stanovsky, Jonathan Bragg, Nicholas Lourie, Jungo Kasai, Yejin Choi, Noah A Smith, and Daniel S Weld. 2021. Genie: A leaderboard for human-in-the-loop evaluation of text generation. *arXiv preprint arXiv:2101.06561* (2021).

Julia Kreutzer, Joshua Uyheng, and Stefan Riezler. 2018. Reliability and learnability of human bandit feedback for sequence-to-sequence reinforcement learning. *arXiv preprint arXiv:1805.10627* (2018).

Antonio L. Lagarda, Daniel Ortiz-Martínez, Vicent Alabau, and Francisco Casacuberta. 2015. Translating without in-domain corpus: Machine translation post-editing with online learning techniques. *Computer Speech & Language* 32, 1 (2015), 109–134. https://doi.org/10.1016/j.csl.2014.10.004 Hybrid Machine Translation: integration of linguistics and statistics.

Tsz Kin Lam, Julia Kreutzer, and Stefan Riezler. 2018. A reinforcement learning approach to interactive-predictive neural machine translation. *arXiv preprint arXiv:1805.01553* (2018).

Tsz Kin Lam, Shigehiko Schamoni, and Stefan Riezler. 2019. Interactive-predictive neural machine translation through reinforcement and imitation. *arXiv preprint arXiv:1907.02326* (2019).

Donghui Lin. 2011. Humans in the Loop of Localization Processes. In *The Language Grid*. Springer, 201–213.

Chien-Liang Liu, Chia-Hoang Lee, Ssu-Han Yu, and Chih-Wei Chen. 2011. Computer assisted writing system. *Expert Systems with Applications* 38, 1 (2011), 804–811.

Elena Lloret, Laura Plaza, and Ahmet Aker. 2013. Analyzing the capabilities of crowdsourcing services for text summarization. *Language resources and evaluation* 47, 2 (2013), 337–369.

Francesco Moramarco, Alex Papadopoulos Korfiatis, Aleksandar Savkov, and Ehud Reiter. 2021. A preliminary study on evaluating Consultation Notes with Post-Editing. *arXiv preprint arXiv:2104.04402* (2021).

Khanh Nguyen, Hal Daumé III, and Jordan Boyd-Graber. 2017. Reinforcement learning for bandit neural machine translation with simulated human feedback. *arXiv preprint arXiv:1707.07402* (2017).

Hiroyuki Osone, Jun-Li Lu, and Yoichi Ochiai. 2021. *BunCho: AI Supported Story Co-Creation via Unsupervised Multitask Learning to Increase Writers' Creativity in Japanese*. Association for Computing Machinery, New York, NY, USA. https://doi.org/10.1145/3411763.3450391

Vishakh Padmakumar and He He. 2021. Machine-in-the-Loop Rewriting for Creative Image Captioning. *arXiv preprint arXiv:2111.04193* (2021).

Tatiana Passali, Alexios Gidiotis, Efstathios Chatzikyriakidis, and Grigorios Tsoumakas. 2021. Towards Human-Centered Summarization: A Case Study on Financial News. In *Proceedings of the First Workshop on Bridging Human–Computer Interaction and Natural Language Processing*. Association for Computational Linguistics, Online, 21–27. https://aclanthology.org/2021.hcinlp-1.4

Álvaro Peris and Francisco Casacuberta. 2018. Active Learning for Interactive Neural Machine Translation
11

# A Appendix

## A.1 Study 1 Method Details: Data Collection, Inclusion Criteria, and Manual Coding Procedures

We collected academic papers by searching three large online indices: Web of Science[10], Engineering Village[11], and ACM Digital Library[12]. Because this is a rapidly growing field and some relevant papers may not be officially published, we also searched arxiv.org[13] for e-print papers. In addition, we included relevant papers that we encountered in previous research activities and snowballed relevant references (for example, we included papers from the workshops of Hum-eval[14] and HCI-NLP[15]).

The specific keywords we used to query the papers include: 1) A word or phrase about AI: AI OR "artificial intelligence" OR machine OR model; 2) A word or phrase related to human-AI collaboration: collaborate OR interact OR assist OR co-author OR co-write OR co-work OR in-the-loop OR co-create OR "human feedback" OR "human-centered"; and 3) A word or phrase about the task: "text summarization" OR "document summarization" OR "article summarization" OR "text generation" OR "document generation" OR "article generation" OR "caption generation" OR "machine translation" OR "style transfer". For each keyword, we searched for all of its forms and/or tenses.

The comprehensive search resulted in 692 papers. After a period of team discussion and iteration, we developed the criteria of the papers that we would like to include in our analysis: First, the paper needed to directly contribute to the problem space of text-to-text generation tasks. Papers about image/video caption generation, speech recognition, and speech-based machine translation were excluded. Second, the paper had to contribute an interface, workflow, or user study that involve human-AI collaboration. For this reason, survey papers were excluded. In addition, because our goal was to identify types of human-AI collaboration that were not obvious previously, we intentionally excluded papers that in which the only human-AI collaboration is human evaluation for model generated final results or human generating training data.

The first author and two other authors then independently coded a random sample of 150 papers out of the 692 collected papers for inclusion or exclusion. Specifically, the first author coded all 150 papers, and the other two researchers coded 75 respectively. They reached a high inter-rater reliability by comparing the last two researchers' coding with the first authors' respectively (average Cohen's Kappa = 0.9), validating the inclusion coding strategy. The first author then coded the remaining 542 papers for inclusion or exclusion. This led to 106 papers that met our inclusion criteria. Due to the rapid advancement in natural language generation in the recent ten years (Dong et al., 2021), we believe insights from recent papers would be most relevant to the design and development of future systems and research. Therefore, we excluded papers published earlier than 2011. All these procedures resulted in a final collection of 70 papers selected for analysis.

Finally, we analyzed the 70 papers following the procedures of thematic analysis (Guest et al., 2011), a manual coding method in qualitative research with which researchers iteratively develop themes from qualitative data. The first author conducted two rounds of open thematic coding on the papers, focusing on aspects such as human actions, collaboration goals, and problem domains. During this process, the team discussed, iterated and refined the themes.

## A.2 Demographic Details of Study 2 Participants

The demographic details of Study 2 participants can be found in Table 2.

---

[10]https://clarivate.com/webofsciencegroup/solutions/web-of-science/
[11]https://www.engineeringvillage.com/
[12]https://dl.acm.org/
[13]https://arxiv.org/
[14]https://humeval.github.io/
[15]https://aclanthology.org/2021.hcinlp-1.0/

| ID | Gender | Age | Occupation | Education | Experience: summarization | Experience: editing | Familiarity: government bills | Familiarity: online news | Familiarity: Reddit posts |
|---|---|---|---|---|---|---|---|---|---|
| P1 | M | 50-59 | Newspaper writer | Bachelor | 6 | 6 | 6 | 6 | 5 |
| P2 | M | 20-29 | Student | Bachelor | 6 | 6 | 6 | 6 | 6 |
| P3 | M | 30-39 | Student | Bachelor | 5 | 7 | 5 | 7 | 7 |
| P4 | F | 60-69 | Freelance editor | Bachelor | 5 | 7 | 5 | 7 | 5 |
| P5 | F | 30-39 | Freelance editor | Doctorate | 7 | 7 | 5 | 7 | 5 |
| P6 | F | 30-39 | Project manager | Master | 6 | 6 | 5 | 7 | 7 |
| P7 | M | 30-39 | Freelancer editor | Bachelor | 5 | 4 | 3 | 6 | 5 |
| P8 | F | 30-39 | Freelance writer | Master | 7 | 7 | 1 | 6 | 1 |
| P9 | F | 30-39 | Marketing consultant | Master | 5 | 7 | 3 | 7 | 7 |
| P10 | F | 20-29 | Student | Bachelor | 7 | 5 | 2 | 6 | 6 |
| P11 | F | 50-59 | Publicist | Bachelor | 7 | 7 | 4 | 7 | 7 |
| P12 | M | 60-69 | Artist | Bachelor | 6 | 6 | 1 | 6 | 1 |
| P13 | F | 30-39 | Freelance writer | Bachelor | 6 | 7 | 2 | 7 | 5 |
| P14 | M | 40-49 | Engineer | Bachelor | 5 | 7 | 1 | 7 | 7 |
| P15 | F | 20-29 | Student | High school | 7 | 7 | 6 | 7 | 7 |
| P16 | F | 30-39 | Student | Bachelor | 7 | 7 | 7 | 7 | 7 |

Table 2: Demographic information of Study 2 participants. All the information are self-reported by the participants. All the participants were based in the United States. Column "Experience: summarization" reports their answers to the question "rate the following statement: 'I am experienced in text summarization' on a scale of 1 to 7, with 1 being least experienced and 7 being most experienced." The other columns on experience or familiarity reports their answers to questions in the same format.

### A.3 Prototype Interfaces Developed for Study 2 Interviews

Figure 3-7 show the screenshots of prototype interfaces explained in §3.2. All interfaces except Writing with Model Assistance contain the same original text (a news article from the articles used in the warm-up activity) that needs to be summarized on the left. The representations of different types of human-AI interactions are on the right side of the interface.

### A.4 Study 2 Additional Findings: User Expectation and Needs in Writing Summaries

**Summarizing with a specific audience in mind.** Many participants had a personal template for writing summaries that they learned in school or work. For example, P6 always checked for *"the who, the what, the where, and the why."* P1 looked for *"the who—Who is it about? What was it about? And then what was the outcome?"* and stated that they used the same strategy in all of the summarization tasks in Part 1.

Participants expressed the need to know the target audience of the summary, so that they could determine what kind of information would be useful to them. As P5 said: *"If I don't know who really is my audience in writing these summaries, I don't know what detail would matter to them."* They would also like to customize the summary to suit the needs of different audiences, especially when the original article was less factual and had more room for interpretation, such as the Reddit posts: *"your audiences is going to determine what's important to put in the summary. An attorney is going to perhaps want different information then your common Joe out on the streets. That affects of how something is summarized, you know, because there's always a choice."* (P12)

**Support on background knowledge.** Participants reported that they faced many challenges in terms of comprehending the original article and writing the summaries. When reading the original articles, many were hindered by the lack of background knowledge. In P10's words, *"it would*

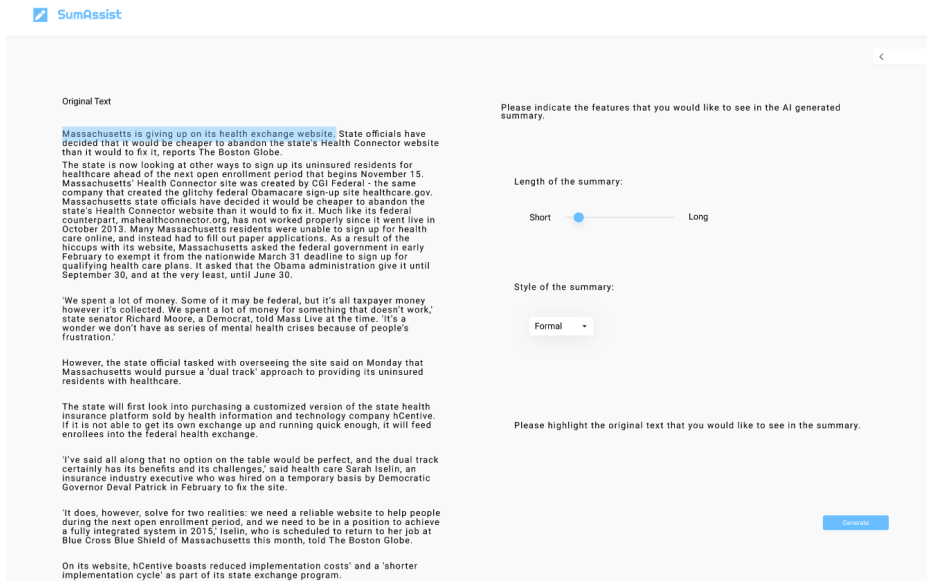

Figure 3: The interface for Guiding Model Output. Users can change the desired summary length and style (formal or informal) using sliders and highlight parts of the original text that they want to include in the summary. Users can press the "Generate" button to get the "AI-generated" summary based on their inputs.

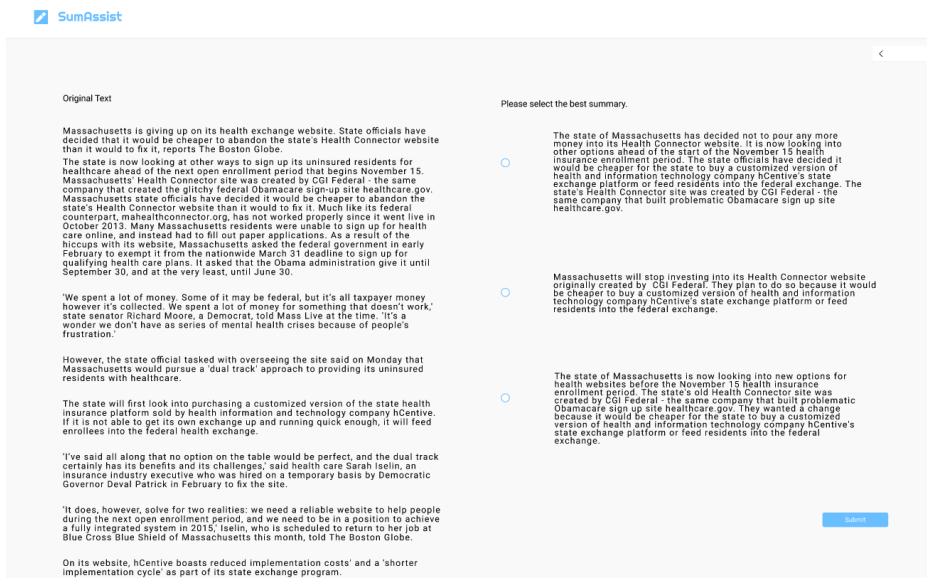

Figure 4: The interface for Selecting or Rating Model Output. Users can chose the final product from three "AI-generated" candidate summaries.

*have helped me to have someone explain what that situation was because I had a very hard time understanding the context behind that situation."* Related to this was the difficulty in understanding jargon specific to a domain that the participants were unfamiliar with. For example, P13 is unfamiliar with the legislative jargon in the Government Bills but considered they *"are really essential for being able to understand kind of the larger overall picture of the text."* P7 shared that the news article about health website contained technical jargon that *"slows down the process and makes it harder."* Therefore, participants hope to get support for the lack of background knowledge. P15 imagined *"an in-text dictionary* support for explaining jargon.

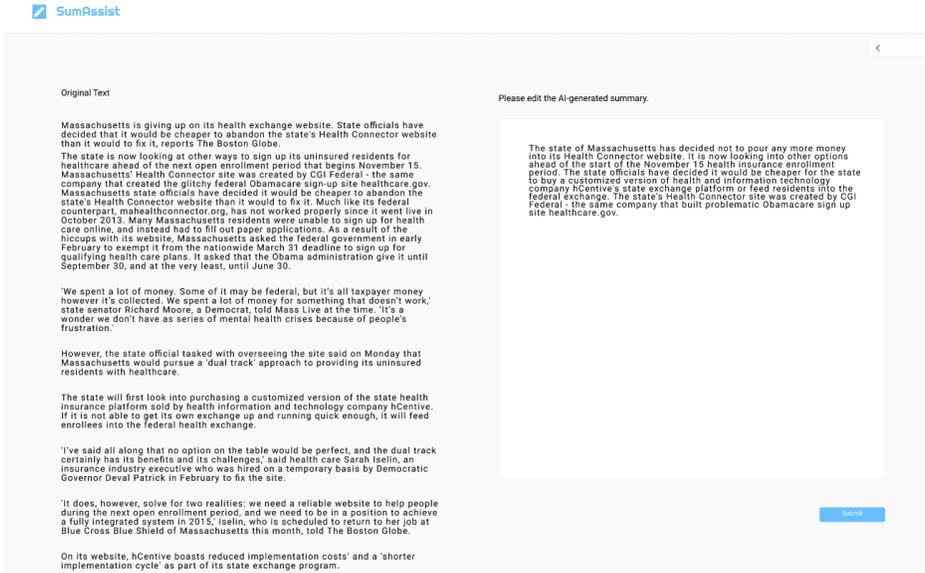

Figure 5: The interface for Post-editing. Users see an "AI-generated" summary in the text box that they can hypothetically edit.

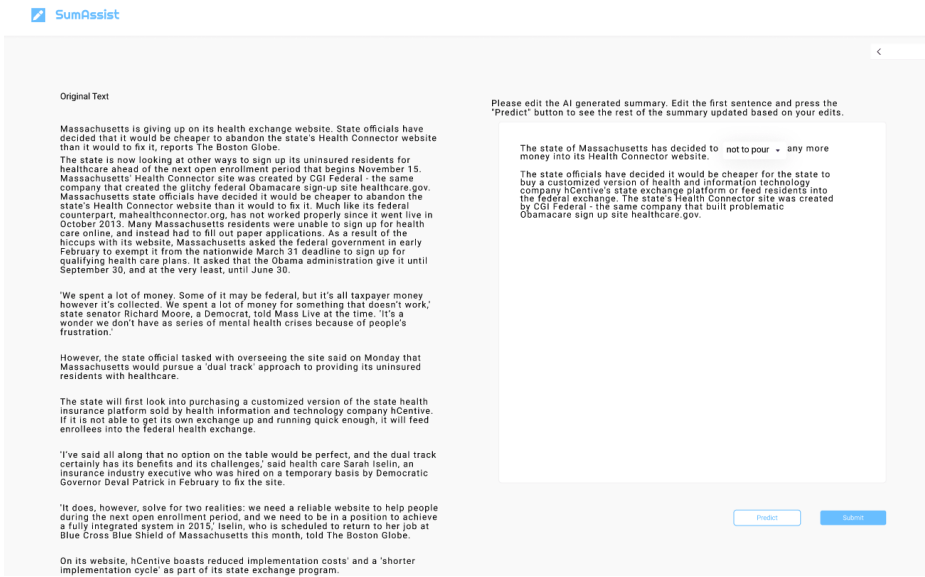

Figure 6: The interface for Interactive Editing. Users see an "AI-generated" summary in the text box. They can use the drop-down menu to change certain words in the first sentence. They can then press "Predict" to request the model to update the rest of the summary based on those edits.

**Summarizing formal and informal texts.** Participants found summarizing different genres of original text challenging in different ways. Participants who thought comprehending the Government Bills challenging described them as *"super dry, super repetitive,"* (P3) and *"not designed as an article."* (P2) It was difficult to identify the crucial information from a Bill since everything seems important: *"they want you to take all of it away.* (P13) On the other hand, although easy to understand, Reddit posts posed a different challenge due to their informal, unfocused style. For example, in P1's words, the Reddit posts were *"more of a narrative... there's no formula to (summarize) them."* Furthermore, many found it challenging to compose the summaries in the same personalized style as the original poster: *"I'm not really familiar with summarizing something in somebody else's voice or in somebody else's tone."* (P3) Contrast to the summarization for the Government Bills, described as having *"only one outcome",* and being *"factual,"* (P7) and *"more direct, more to the point,"* (P10) Reddit posts come with *"an element of opinion to it,"* (P7) and therefore could be summarized through different angles. Some thought they should write in the same perspective as the original poster and preserve the opinions, as P9 said, *"I would make sure I picked up what they were trying to, to lay out."* others considered a summary as a neutral representation of information and thus they had to *"pull the opinions out and give the general message."* (P14)

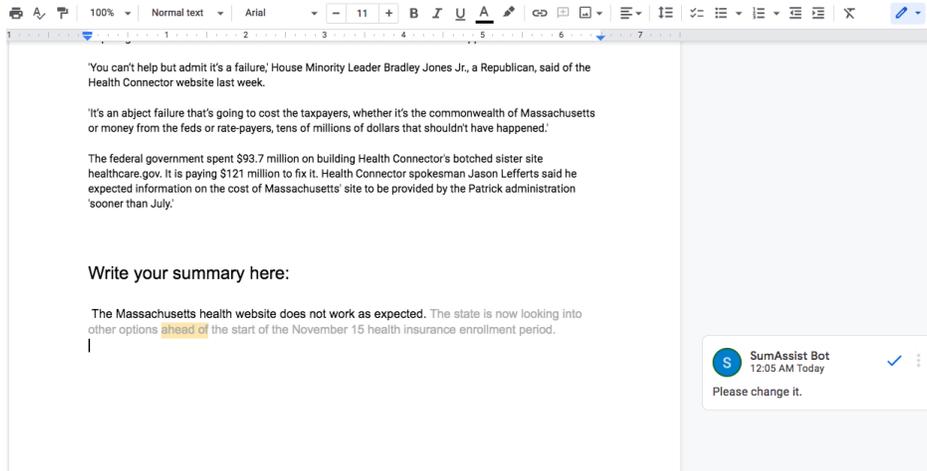

Figure 7: The interface for Writing with Model Assistance. In a Google Doc, users can see the original article on the top and they can write their summary under the section "Write your summary here:". First, the user types a sentence for their summary, then a Bot (played by a researcher who log in with the "SumAssist Bot account") will insert the next sentence in gray fonts. The Bot will also insert comments on words in the user written sentence and suggest them to make changes.

### A.5 Interview protocol and interaction instruction.

Please see the interview protocol and interaction instruction on the next page. Since we need to protect the privacy of our participants, we are not able to share the full interview transcripts.

# A.5 Interview Protocol

**(§3.3 contains interaction instruction and questions for each of the five prototypes)**

## 1. Overview

First of all thank you for spending the time and talking with me today. We are researchers studying how we can help people summarize text. Your participation in this study could help us understand this topic, and contribute to scientific knowledge - ultimately we may publish a research paper about insights from your participation in our study.

## 2. Consent

This interview will last about 90 minutes. I would like to talk to you about your experience with text summarization. You should have already finished the warm-up for the study, and I will ask you questions about it. I will also have you try out some new tools that we developed to support text summarization and ask you questions along the way.

There are no right or wrong answers to my questions. I want to hear your personal experiences and opinions. You also absolutely can decline to answer any of the questions that I ask. At any point if you feel uncomfortable or need to pause or take a break, just let me know.

Everything that is said in this session will be completely confidential and used for research purposes only. Your responses may be published as quotes in a future academic research paper, but no identifiers will tie your username or any other identifiable factors to those responses.

For note-taking purposes, I will record the audio of this call. I will ask you to share your screen while you are trying out our new tools, and I will also record your screen. Feel free to turn off your camera if you don't like your image to be recorded.

Any questions before we start?

Is it okay if I start studying now?

[if they say yes, start recording]

## 3. Interview

### 3.1 Opening

To start, could you please tell me a little bit about yourself?

- What experience do you have with writing and editing?

3.2 Reflection on warm-up

Could you please walk me through your experience writing the summaries?
- How did you approach the tasks?
- What were some strategies that you took?
    - How did you decide what to include?
    - How did you decide if a summary is done and good to go?
- What went well?
- What did not go well?
- What were some challenges?
- How do you feel about summarizing the different articles?
    - Which ones were easy? Why
    - Which ones were difficult? Why?
    - How did you approach the different articles differently?

[Pick a the task that they thought as most difficult]
Could you tell me more about your experience finishing this task? Walk me through how you did it.
- Why was this task particularly difficult?
- Did you approach this task differently than other ones?
    - If yes, how? And why?
    - If not, how did your strategy with previous articles work or not work?
- What went well?
- What did not go well?
- What are some challenges?
- What kind of support do you wish to have?
    - Do you want to have the same support for the other tasks? Why or why not?

Did you take any notes or use any external tools or resources?

In general, what kind of support do you wish you could have while summarizing articles?
- Do you want to have different support for different types of articles? Why or why not?

3.3 Interaction with prototypes

Now I am going to have you try out SumAssists, which is a collection of digital tools that we developed to support people to write summaries. I will ask you questions as you try out the tools. Note that some features of the tools are not fully developed yet, so for some parts I need to ask you to imagine your interaction with some features.

I'm now asking you to share your screen to show me your interaction with the tools.

[send them the link to the Figma prototype]

Now you are seeing the original article. This is the same news article about the Massachusetts heath website that you have worked on in Part 1 of the study. You can see 5 different tools that are designed to support you to write the summary for this article.

*3.3.1 Selecting or rating model output*

[ask the participant to interact with the tool]

What's your first impression of this tool?
- What do you think it is for?
- How do you want to use it?

This tool provides you three options of AI-generated summaries of this article. You will be able to select the one that you think is the best.

Now please imagine that you are summarizing the article using this tool. Please talk through step-by-step how you are going to do it.
- What do you like about this tool? Why?
- What do you dislike about this tool? Why?
- What is helpful with your summarization process? In what ways?
- What is unhelpful with your summarization process? In what ways?
- How much power of control with the summarization do you feel that you have using this tool? Do you like it this way? Why?
- How much will you rely on this tool while doing the summarization task? Do you like it this way? Why?
  - What type of documents will you rely on this tool
  - How much would you use it
  - Any cases that you wouldn't want to use it
  - What case you want to totally override it
- How much will you trust the AI-generated summaries using this tool? Do you like it this way? Why?

If we are making this tool available in real life, in what situation do you see yourself using this tool?

Imagining you are summarizing the Reddit posts using this tool, how would your experience be different or similar?

Imagining you are summarizing the government bills using this tool, how would your experience be different or similar?

How would you like to improve this tool?

Let's go back to the main page by clicking on the arrow on the upper right corner.

*3.3.2 Post-editing*

[ask the participant to interact with the tool]

What's your first impression of this tool?
- What do you think it is for?
- How do you want to use it?

[Debrief them about the tool] This tool provides you with an AI-generated summaries of this article that you can edit on. In the real world you will be able to type and delete any words in this text box on the right just as you are doing normal editing. Right now our implementation doesn't support that. Just imagine you can type and edit the summary.

Now please imagine that you are summarizing the article using this tool. Please talk through step-by-step how you are going to do it.
- What do you like about this tool? Why?
- What do you dislike about this tool? Why?
- What is helpful with your summarization process? In what ways?
- What is unhelpful with your summarization process? In what ways?
- How much power of control with the summarization do you feel that you have using this tool? Do you like it this way? Why?
- How much will you rely on this tool while doing the summarization task? Do you like it this way? Why?
  - What type of documents will you rely on this tool
  - How much would you use it
  - Any cases that you wouldn't want to use it
  - What case you want to totally override it
- Compared to the previous tool, which one do you think would be more helpful? Why?

If we are making this tool available in real life, in what situation do you see yourself using this tool?
- Imagining you are summarizing the Reddit posts using this tool, how would your experience be different or similar?
  - Compared to the previous tool, how do you think this tool would be more or less helpful in this situation?

- Imagining you are summarizing the government bills using this tool, how would your experience be different or similar?
    - Compared to the previous tool, how do you think this tool would be more or less helpful in this situation?
    - If the AI is doing a reasonable job, for longer docs that you may not know as much
    - Are there other text that this tool will be helpful for you to summarize

How would you like to improve this tool?

Let's go back to the main page

*3.3.3 Interactive editing*

[ask the participant to interact with the tool]

What's your first impression of this tool?
- What do you think it is for?
- How do you want to use it?

[Debrief them about the tool] This tool provides you with an AI-generated summaries of this article. You will be able to edit the first sentence of the summary by selecting in the drop down menu. Once you make the selection and click predict, the rest of the summary will be updated based on your edits on the first sentence.

Now please imagine that you are summarizing the article using this tool. Please talk through step-by-step how you are going to do it.
- What do you like about this tool? Why?
- What do you dislike about this tool? Why?
- What is helpful with your summarization process? In what ways?
- What is unhelpful with your summarization process? In what ways?
- How much power of control with the summarization do you feel that you have using this tool? Do you like it this way? Why?
- How much will you rely on this tool while doing the summarization task? Do you like it this way? Why?
- How much will you trust the AI-generated summaries using this tool? Do you like it this way? Why?
- Compared to the previous tool, which one do you think would be more helpful? Why?

If we are making this tool available in real life, in what situation do you see yourself using this tool?
- Imagining you are summarizing the Reddit posts using this tool, how would your experience be different or similar?

- Compared to the previous tool, how do you think this tool would be more or less helpful in this situation?
- Imagining you are summarizing the government bills using this tool, how would your experience be different or similar?
    - Compared to the previous tool, how do you think this tool would be more or less helpful in this situation?

How would you like to improve this tool?

Let's go back to the main page

### 3.3.4 Guiding model output

[ask the participant to interact with the tool]

What's your first impression of this tool?
- What do you think it is for?
- How do you want to use it?

[Debrief them about the tool] This tool provides you the power to tell the AI model to generate the kind of summaries that you'd like to see. You can change the length of the summary, the style of the summary, and also highlight parts of the original sentences that you'd like to see in the summary by highlighting them in the original text.

Now please imagine that you are summarizing the article using this tool. Please talk through step-by-step how you are going to do it.
- What do you like about this tool? Why?
- What do you dislike about this tool? Why?
- What is helpful with your summarization process? In what ways?
- What is unhelpful with your summarization process? In what ways?
- How much power of control with the summarization do you feel that you have using this tool? Do you like it this way? Why?
- How much will you rely on this tool while doing the summarization task? Do you like it this way? Why?
- How much will you trust the AI-generated summaries using this tool? Do you like it this way? Why?
- Compared to the previous tool, which one do you think would be more helpful? Why?

If we are making this tool available in real life, in what situation do you see yourself using this tool?

- Imagining you are summarizing the Reddit posts using this tool, how would your experience be different or similar?
    - Compared to the previous tool, how do you think this tool would be more or less helpful in this situation?
- Imagining you are summarizing the government bills using this tool, how would your experience be different or similar?
    - Compared to the previous tool, how do you think this tool would be more or less helpful in this situation?

How would you like to improve this tool?

Let's go back to the main page

*3.3.5 Writing with model assistance*

[ask the participant to interact with the tool]

Now this tool is embedded in the Google doc. For this one, please type the summary that you'd like to write for this article.
[Participant type summary]

[SumAssist Bot copies a sentence in gray color once they finish the first sentence]
[SumAssist Bot adds a comment on their first sentence]

If you think the sentence in gray is good, go ahead and keep writing. If you want to edit it, also go ahead. If you don't like it, you can just delete it and write.

[SumAssist Bot copies a sentence in gray color once they finish the first sentence]
[SumAssist Bot adds a comment on their first sentence]
[ask the participant to stop]

This tool is a bot that we built in Google doc and it can suggest next sentences in gray text. It can also comment on the sentence that you wrote as you are writing.

Imagine that you are writing the entire summary with this bot. What do you think the experience would be like?
- What is helpful with your summarization process? In what ways?
- What is unhelpful with your summarization process? In what ways?
- What do you like about this tool? Why?
- What do you dislike about this tool? Why?
- How do you feel about the auto suggestions?

- How could the auto suggestion be more useful?
- How do you feel about the auto commenting?
  - How could the auto commenting be more useful?
- Would the auto suggestions and comments be annoying? Why and why not?
- How much power of control with the summarization do you feel that you have using this tool? Do you like it this way? Why?
- How much will you rely on this tool while doing the summarization task? Do you like it this way? Why?
- How much will you trust the AI-generated summaries using this tool? Do you like it this way? Why?
- Compared to the previous tool, which one do you think would be more helpful? Why?

If we are making this tool available in real life, in what situation do you see yourself using this tool?
- Imagining you are summarizing the Reddit posts using this tool, how would your experience be different or similar?
  - Compared to the previous tool, how do you think this tool would be more or less helpful in this situation?
- Imagining you are summarizing the government bills using this tool, how would your experience be different or similar?
  - Compared to the previous tool, how do you think this tool would be more or less helpful in this situation?

How would you like to improve this tool?

3.4 General Reflection

In general, which tools that you would most likely be using when you are summarizing articles? Why?

In general, which tools that you would least likely be using when you are summarizing articles? Why?
- For Reddit, news, and government bills respectively?

If you could snap your fingers and create a summarization assistant to help you, how would you like to combine some of these tools?
- And what other features would you like to have?